\documentclass[sigconf]{acmart}
\AtBeginDocument{%
  }

\copyrightyear{2027}
\acmYear{2027}
\setcopyright{cc}
\setcctype{by}
\acmConference[WSDM '27]{Proceedings of the Twentieth ACM International Conference on Web Search and Data Mining}{February 15--19, 2027}{Hong Kong, Hong Kong}
\acmBooktitle{Proceedings of the Twentieth ACM International Conference on Web Search and Data Mining (WSDM '27), February 15--19, 2027, Hong Kong, Hong Kong}
\acmDOI{10.1145/3862322.3863901}
\acmISBN{979-8-4007-3117-4/2027/02}

\usepackage{tabularx}
\usepackage{array}

\newcolumntype{Y}{>{\centering\arraybackslash}X}

\usepackage{booktabs}
\usepackage[table]{xcolor}

\usepackage{enumitem}
\usepackage{multirow}

\usepackage[most]{tcolorbox}

\newtcolorbox{skillbox}[1]{
    enhanced,
    breakable,
    colback=gray!4,
    colframe=gray!35,
    boxrule=0.5pt,
    arc=1.5pt,
    left=4pt,
    right=4pt,
    top=3pt,
    bottom=3pt,
    before skip=3pt,
    after skip=4pt,
    title=\textbf{#1},
    coltitle=black,
    colbacktitle=gray!10,
    fonttitle=\small,
    toptitle=2pt,
    bottomtitle=2pt
}

\newtcolorbox{promptbox}[1]{
    enhanced,
    breakable,
    colback=white,
    colframe=gray!35,
    boxrule=0.5pt,
    arc=1.5pt,
    left=5pt,
    right=5pt,
    top=3pt,
    bottom=3pt,
    before skip=3pt,
    after skip=4pt,
    title=\textbf{#1},
    coltitle=black,
    colbacktitle=gray!7,
    fonttitle=\small,
    toptitle=2pt,
    bottomtitle=2pt
}
\begin{document}

\title{CRAFTS: Candidate-Ranking Agent with Feedback-Refined Transferable Skills}

\author{Jiayin Feng}
\email{fengjiayin@stu.cqu.edu.cn}
\orcid{0009-0000-2199-679X}
\affiliation{%
  \institution{Chongqing University}
  \city{Chongqing}
  \country{China}
}

\author{Min Gao}
\authornote{Corresponding authors.}
\email{gaomin@cqu.edu.cn}
\orcid{0000-0003-0127-7477}
\affiliation{%
  \institution{Chongqing University}
  \city{Chongqing}
  \country{China}
}

\author{Zongwei Wang}
\email{zongwei@cqu.edu.cn}
\orcid{0009-0003-2606-9508}
\affiliation{%
  \institution{Chongqing University}
  \city{Chongqing}
  \country{China}
}

\author{Bincheng Gu}
\email{gubincheng@stu.cqu.edu.cn}
\orcid{0009-0006-1734-299X}
\affiliation{%
  \institution{Chongqing University}
  \city{Chongqing}
  \country{China}
}
\author{Junliang Yu}
\authornotemark[1]
\email{junliang.yu@griffith.edu.au}
\orcid{0000-0003-3401-9829}
\affiliation{%
  \institution{Griffith University}
  \city{Brisbane}
  \country{Australia}
}


\renewcommand{\shortauthors}{Feng et al.}
\begin{abstract}

LLM-based recommendation agents have demonstrated considerable potential in understanding user preferences and supporting personalized decision-making. However, existing approaches primarily rely on instance-specific interaction contexts or case-level external memory for preference modeling. While useful, such information is often redundant and context-bound, providing limited reusable guidance for preference interpretation and candidate comparison. To address this limitation, we propose CRAFTS, a framework that abstracts recurring patterns of preference judgment and candidate comparison from historical recommendation experiences into transferable decision skills. Given a new recommendation instance, CRAFTS retrieves the skills most relevant to its interaction context. To bridge the gap between skill retrieval and effective skill use, CRAFTS introduces skill-guided trajectory supervision. Specifically, a high-capacity teacher model generates recommendation trajectories that demonstrate when and how to apply the retrieved skills. These trajectories supervise a lightweight recommendation agent to selectively integrate relevant skills into candidate-level ranking decisions. CRAFTS further performs feedback-driven Skill Bank evolution by analyzing failed recommendation trajectories to identify missing or insufficient decision patterns and update the Skill Bank accordingly. Experiments on multiple recommendation benchmarks demonstrate the effectiveness of CRAFTS while ablation studies further validate the effectiveness of transferable skill learning and adaptive skill utilization. The source code is available at \url{https://github.com/ooYooYoo/CRAFTS_code}.

\end{abstract}




\maketitle

\section{Introduction}
\begin{figure}[!t]
    \centering
    \includegraphics[width=\columnwidth]{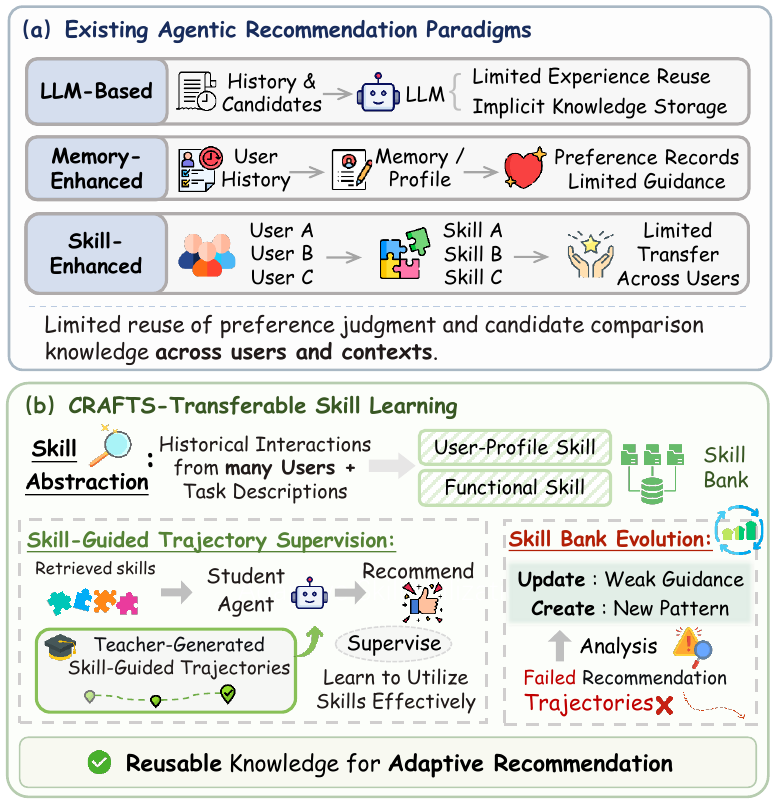}
    \caption{Comparison of (a) existing agentic recommendation paradigms and (b) CRAFTS, which learns and reuses transferable skills for adaptive recommendation.}
    \Description{A comparison of existing agentic recommendation paradigms that rely on instance-specific contexts or external memory and CRAFTS, which learns reusable recommendation skills and transfers them across users and recommendation contexts.}
    \label{fig:intro}
\end{figure}
Recommender systems aim to identify items that match users' interests from a large candidate space, thereby supporting personalized information access and decision-making. With recent advances in large language models (LLMs), LLM-based recommendation has introduced more flexible language-driven reasoning and interaction into the recommendation process~\cite{wu2024survey,geng2022recommendation}. Building on this trend, recommendation has evolved from static preference prediction toward more interactive and autonomous decision processes, leading to the emergence of agentic recommender systems, in which recommendation agents can reason over user histories, interact with users, and adapt their decisions over time~\cite{shi2024large,wang2024recmind,wang2024macrec,cai2025agentic}. In a typical agentic recommender system, a recommendation agent receives the current user history and candidate items, analyzes user preferences, compares candidate items, and produces a recommendation decision. When interaction or feedback is available, the agent can further adjust its reasoning and subsequent recommendations.

To support the continual improvement of agentic recommender systems, existing methods have begun to externalize recommendation experience as knowledge units that can be independently maintained and updated. Memory-enhanced methods, for example, store users' past interactions, personalized profiles, or collaborative signals in external memory and retrieve relevant records to support current decisions~\cite{xu2025iagent,chen2026memrec}. However, such memories mainly preserve factual evidence about what users have preferred in the past, while providing limited procedural guidance on how this evidence should be translated into comparisons among candidate items and ranking decisions. Recent skill-enhanced methods take a step further, equipping each user with a dedicated policy skill that encodes personalized decision principles, allowing the agent to adapt how it reasons rather than only what it remembers~\cite{tao2026sager}. Nevertheless, this design remains strongly user-specific: each user's skill is learned and evolved primarily from that user's own interaction history, so the decision knowledge available to the agent is bounded by the coverage and diversity of a single user's experience. In practice, users with similar preferences may encounter different contexts and exhibit complementary behaviors. Meanwhile, existing recommendation studies have shown that cross-user signals can provide complementary evidence for inferring target users' preferences, helping identify latent preferences that cannot be sufficiently inferred from their individual interaction histories alone~\cite{chen2026memrec,liu2025multi}. This suggests that recurring decision patterns revealed across preference-similar users can be abstracted into transferable recommendation skills and organized in a shared knowledge space, allowing reusable decision knowledge to be jointly maintained and utilized.

Based on these insights, we propose CRAFTS (\textbf{C}andidate-\textbf{R}anking \textbf{A}gent with \textbf{F}eedback-Refined \textbf{T}ransferable \textbf{S}kills), a framework for learning, utilizing, and continually refining transferable recommendation skills in agentic recommender systems. At its core, CRAFTS maintains a transferable skill bank that serves as a shared knowledge space across users and recommendation contexts. CRAFTS constructs this bank by abstracting reusable preference patterns from historical interactions into user-profile skills and organizing general candidate-comparison strategies as functional skills. For each recommendation instance, context-relevant skills are retrieved and selected from the skill bank to support preference analysis and candidate ranking. To continually improve the transferable knowledge space, CRAFTS analyzes failed recommendation trajectories to identify missing or insufficient decision patterns and uses this feedback to refine existing skills or introduce new ones~\cite{shinn2023reflexion,madaan2023self,zhang2026memskill}. In this way, fragmented recommendation experience can be progressively transformed into reusable decision knowledge that is accumulated and transferred across users and recommendation contexts.

Although the shared Skill Bank makes recommendation skills reusable and transferable, retrieving relevant skills does not guarantee their effective utilization. Transferable skills encode abstract decision principles distilled from recurring patterns across users and contexts, and therefore a retrieved skill may only partially match the current user's history and candidate set. The recommendation agent should still determine which skill signals are applicable, ground them in instance-specific preference evidence, and translate them into concrete comparisons among candidate items. Simply providing retrieved skills as additional context leaves this skill-application process implicit and offers no explicit supervision on how such principles should be instantiated for the current recommendation decision.

To make this skill-application process learnable, CRAFTS introduces skill-guided trajectory supervision. Given the current recommendation context and selected skills, a teacher LLM is used to externalize the otherwise implicit process into structured recommendation trajectories. These trajectories explicitly demonstrate which skill signals should be used, how they should be grounded in the current user context, and how they should guide candidate comparison and ranking~\cite{gao2025process}. The recommendation agent is then trained on these trajectories to learn how to selectively apply transferable skills and translate them into candidate-level ranking decisions. Through this process, CRAFTS moves beyond treating transferable skills as additional external context and enables the recommendation agent to internalize their adaptive utilization as a stable recommendation capability.

The main contributions of this paper are summarized as follows:
\begin{itemize}[leftmargin=*, itemsep=0pt, topsep=2pt]

\item We propose CRAFTS, a framework that abstracts recurring decision patterns from historical interactions into transferable recommendation skills and refines them using feedback from failed trajectories for reuse across users and contexts.
\item We introduce skill-guided trajectory supervision, in which trajectories generated by a teacher LLM train a recommendation agent to adapt retrieved skills to the current context and produce candidate rankings.
\item Experiments on multiple recommendation datasets demonstrate the effectiveness of CRAFTS compared with representative agent-based and skill-enhanced baselines, while ablation studies validate the contributions of its key components.

\end{itemize}

\section{Related Work}
\subsection{LLM-Based Recommendation}

With the improvement of LLM reasoning capabilities, their applications in recommender systems have been increasingly explored~\cite{bao2024large}. Existing studies can be broadly categorized into two lines. The first treats LLMs as semantic enhancement tools, generating item descriptions, user profiles, or external knowledge to enrich traditional recommendation models~\cite{xi2024towards}. The second formulates recommendation as a language modeling task, where user histories and candidate items are organized through prompts to enable preference understanding and candidate selection~\cite{lin2024bridging,he2023large}. Based on this paradigm, some studies fine-tune LLMs on recommendation data to acquire task-specific knowledge~\cite{bao2023tallrec,liao2024llara,zhang2025collm}. For example, TALLRec aligns LLMs with recommendation tasks through instruction tuning~\cite{bao2023tallrec}, while LLaRA maps behavioral representations learned by sequential recommenders into the LLM input space and combines them with item text information for sequential recommendation~\cite{liao2024llara}.

The introduction of recommendation agents further formulates recommendation as a flexible decision-making process~\cite{shi2024large}. RecMind improves the reasoning capabilities of recommendation agents by incorporating self-inspiration mechanisms and external tool utilization into the recommendation process~\cite{wang2024recmind}. MACRec introduces a multi-agent collaborative framework for recommendation by assigning user preference analysis, candidate item analysis, and final recommendation decision-making to different specialized agent roles~\cite{wang2024macrec}. RAH places an LLM agent between users and recommender systems to support human-centered recommendation~\cite{shu2024rah}. Agent4Rec, RecAgent, and AgentCF model user simulation or user--item interactions with LLM-based agents~\cite{zhang2024generative,wang2025user,zhang2024agentcf}. AFL further establishes an interactive feedback loop between the recommendation agent and the user agent, enabling the recommendation agent to adaptively adjust its subsequent recommendation decisions based on the feedback provided by the user agent~\cite{cai2025agentic}. CoARS leverages interaction trajectories and self-distilled reinforcement learning to enable the co-evolution of recommender and user agents~\cite{wang2026self}. However, existing LLM-based recommendation agents mainly focus on instance-level reasoning and interaction, while lacking mechanisms for learning and reusing transferable recommendation skills across scenarios.

\subsection{Memory and Skill Enhancement}

Memory mechanisms enable LLM agents to support long-term interaction and personalized modeling. Existing studies store interaction histories, task states, or experiences in external memory modules and retrieve relevant memories for subsequent decision-making~\cite{packer2023memgpt,zhong2024memorybank,xu2026mem}. In recommendation scenarios, memory mechanisms update user profiles or propagate cross-user signals to alleviate sparse individual histories~\cite{xu2025iagent,tang2025interactive,chen2026memrec}. Specifically, i$^2$Agent introduces dynamic memory that continuously learns from user feedback and updates personalized profiles~\cite{xu2025iagent}, while MemRec constructs a collaborative memory graph to propagate cross-user semantic signals and alleviate sparsity~\cite{chen2026memrec}. These methods enhance the ability to memorize and utilize user preferences. However, they mainly focus on recording ``what preferences users have shown'', while recommendation also requires knowing how to compare and rank candidates based on these preferences.

Unlike memories that emphasize information storage, skills focus on callable operation patterns or decision principles~\cite{zhao2024expel,wang2023voyager}. MemSkill abstracts fixed memory management operations into learnable and evolvable memory skills, forming a closed-loop optimization process through skill selection, execution, and hard-sample-driven evolution~\cite{zhang2026memskill}. SAGER introduces skills into recommendation agents by maintaining self-evolving policy skills for users, enabling personalized decision principles beyond memory content~\cite{tao2026sager}. AgentCom further introduces a shared communication skill bank for agentic recommender systems, enabling personalized skill routing and failure-driven skill evolution~\cite{wang2026personalized}. Unlike SAGER's user-specific policy skills and AgentCom's communication skills, our work learns transferable recommendation skills by organizing reusable preference and candidate comparison patterns into an evolvable Skill Bank. Recommendation trajectories are further leveraged to supervise adaptive skill utilization and refine the learned skill space.

\section{Method}

\begin{figure*}[!t]
    \centering
    \includegraphics[width=0.95\textwidth]{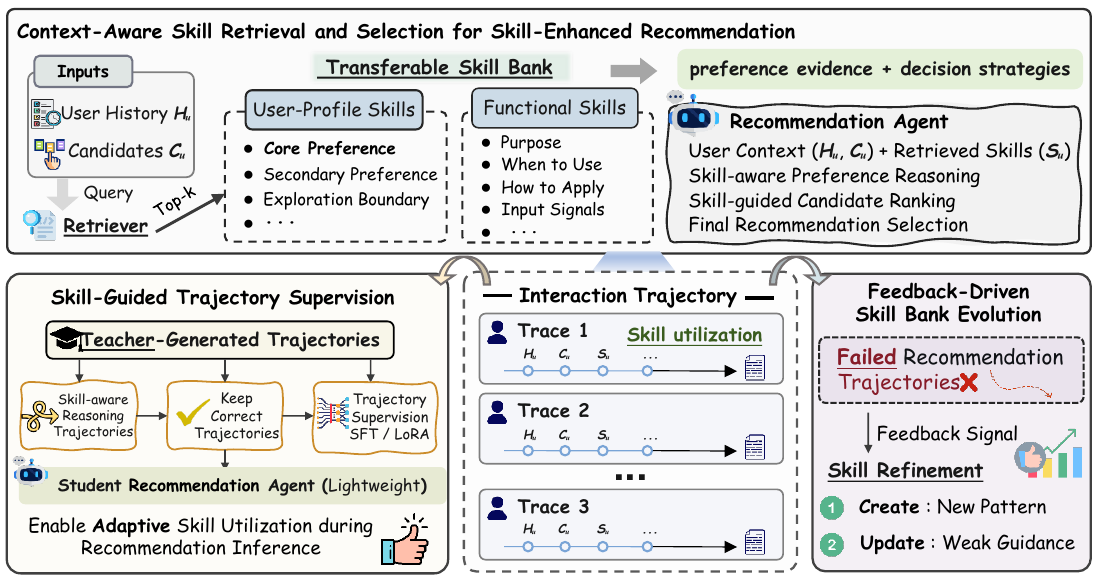}
    \caption{Overview of the proposed CRAFTS framework.}
    \Description{The CRAFTS framework consists of Skill Bank construction, skill retrieval and selection, skill-guided trajectory supervision, and feedback-driven Skill Bank evolution. User-profile and functional skills guide recommendation decisions, while failed recommendation trajectories provide feedback for refining the Skill Bank.}
    \label{fig:framework}
\end{figure*}

In this section, we present \textbf{CRAFTS}, a candidate-ranking agent that learns transferable recommendation skills and their adaptive utilization. CRAFTS constructs a shared Skill Bank and trains the recommendation agent with skill-guided trajectories. For each recommendation instance, context-relevant skills are retrieved to guide candidate ranking. Failed recommendation trajectories are further used to refine the Skill Bank. The overall framework is illustrated in Figure~\ref{fig:framework}.

\subsection{Problem Formulation}

Given a user $u$, let $\mathcal{H}_u=(i_1,i_2,\ldots,i_t)$ denote the user's historical interaction sequence, where $i_t$ denotes the item interacted with by user $u$ at the $t$-th time step. Let $\mathcal{C}_u=\{c_1,c_2,\ldots,c_n\}$ denote the candidate item set. The goal of recommendation is to select the item that the user is most likely to interact with from $\mathcal{C}_u$.

For each recommendation instance, CRAFTS selects an instance-specific skill set $\mathcal{S}_u$ from the Skill Bank. Given the historical interaction sequence, candidate item set, and selected skill set, the recommendation objective is formulated as:
\begin{equation}
\widehat{i}_u = \arg\max_{c \in \mathcal{C}_u} f_{\theta}(c \mid \mathcal{H}_u, \mathcal{C}_u, \mathcal{S}_u),
\end{equation}
where $f_{\theta}(\cdot)$ denotes the preference estimation function of the recommendation agent under the skill-enhanced context, and $\mathcal{S}_u$ denotes the transferable skill set utilized for the current recommendation instance.

\subsection{Skill Bank Construction}

The Skill Bank organizes transferable recommendation knowledge. It contains two types of skills: \textit{user-profile skills}, which capture reusable preference patterns derived from user histories, and \textit{functional skills}, which encode general candidate-comparison and ranking strategies. To account for domain-specific preference semantics, CRAFTS maintains a separate Skill Bank for each domain $d$:
\begin{equation}
\mathcal{B}^{d}=\mathcal{B}_{p}^{d}\cup\mathcal{B}_{f}^{d},
\end{equation}
where $\mathcal{B}_{p}^{d}$ and $\mathcal{B}_{f}^{d}$ denote the user-profile and functional skill sets, respectively.

\textbf{Skill initialization.}
We initialize the two types of skills from complementary sources using an LLM. For functional skills, CRAFTS prompts the LLM with general recommendation principles and task descriptions to generate structured decision strategies. For each training user $u$, we prompt the LLM with the user's interaction history $\mathcal{H}_u$ to generate a structured user profile $p_u$ that captures abstract and reusable preference patterns rather than copying specific interacted items. These user profiles are subsequently aggregated into transferable user-profile skills.

\textbf{User-profile skill aggregation.}
To avoid maintaining fragmented user-specific skills, we further aggregate preference-similar user profiles. Each profile $p_u$ is encoded into a semantic representation $\mathbf{z}_u=E(p_u)$ using a frozen sentence encoder, and profile similarity is measured by cosine similarity:
\begin{equation}
\mathrm{sim}(u,v)=
\frac{\mathbf{z}_u^{\top}\mathbf{z}_v}
{\|\mathbf{z}_u\|_2\|\mathbf{z}_v\|_2}.
\end{equation}
Profiles whose similarity exceeds a predefined threshold are grouped together. Within each cluster, CRAFTS collects the core and secondary preference signals from member profiles and identifies recurring patterns according to their occurrence across users. These dominant patterns are then consolidated into a group-level user-profile skill, reducing user-specific noise and improving skill reusability across similar users. The resulting user-profile skills, together with the initialized functional skills, constitute the initial Skill Bank, which provides transferable recommendation knowledge for subsequent skill selection and recommendation generation.

\subsection{Skill Retrieval and Selection}

Given a recommendation context $(\mathcal{H}_u,\mathcal{C}_u)$, CRAFTS selects an instance-specific skill set $\mathcal{S}_u$ from the current Skill Bank $\mathcal{B}^{d}$. User-profile skills are retrieved according to their relevance to the current context, whereas functional skills are selected by matching the context against their predefined applicability conditions.

\textbf{User-profile skill retrieval.} The user-profile skill set $\mathcal{B}_{p}^{d}$ contains diverse preference patterns, only a small subset of which is relevant to a given recommendation instance. We therefore train a task-specific retriever that uses a preference query $q_u$ to identify the most relevant user-profile skills from $\mathcal{B}_{p}^{d}$.

To train the retriever, we construct supervision data from successful training trajectories generated with the initial semantic-matching procedure. For each trajectory, the user-profile skills selected for the corresponding recommendation instance are treated as positive skills $s^{+}$, while negative skills $s^{-}$ are sampled from the remaining user-profile skills in $\mathcal{B}_{p}^{d}$. The retriever training set is defined as:
\begin{equation}
\mathcal{D}_{r} = \{(q_u, s^{+}, s^{-})\},
\end{equation}
where $\mathcal{D}_{r}$ denotes the retriever training set, and $q_u$ summarizes the preference clues contained in the current user history and candidate set. 

For user-profile skill retriever modeling, we adopt a frozen text encoder $E(\cdot)$ to obtain the basic semantic representations of queries and skills, and then use a trainable projection layer $h_{\psi}(\cdot)$ to map them into a task-specific retrieval space~\cite{karpukhin2020dense}. The relevance score between query $q_u$ and user-profile skill $s$ is defined as:
\begin{equation}
r_{\psi}(q_u, s) = h_{\psi}(E(q_u))^{\top} h_{\psi}(E(s)),
\end{equation}
where $\psi$ denotes the trainable parameters of the projection layer, and $r_{\psi}(q_u,s)$ measures the relevance between query $q_u$ and skill $s$ in the learned retrieval space.

We train the retriever with a pairwise ranking objective~\cite{rendle2012bpr}:
\begin{equation}
\mathcal{L}_{\mathrm{retr}} = - \sum_{(q_u, s^{+}, s^{-}) \in \mathcal{D}_{r}} \log \sigma \left( r_{\psi}(q_u, s^{+}) - r_{\psi}(q_u, s^{-}) \right),
\end{equation}
where $\sigma(\cdot)$ denotes the sigmoid function. This objective encourages the retriever to assign higher relevance scores to positive user-profile skills than to negative ones for the same query, thereby learning a task-specific skill retrieval strategy.

After training, the system retrieves the top-$K$ user-profile skills from the corresponding skill set based on $r_{\psi}(q_u, s)$:
\begin{equation}
\mathcal{S}_{u,p}
=
\operatorname*{arg\,top\text{-}K}_{s \in \mathcal{B}_{p}^{d}}
r_{\psi}(q_u, s),
\end{equation}
where $\mathcal{B}_{p}^{d}$ denotes the user-profile skill set in domain $d$, $K$ denotes the number of retrieved user-profile skills, and $\mathcal{S}_{u,p}$ denotes the resulting set of retrieved user-profile skills. In our implementation, we set $K=2$.

\textbf{Functional skill selection.} Each functional skill is associated with predefined applicability conditions and input signals. CRAFTS matches the current recommendation context against these conditions and selects up to two applicable functional skills, denoted as $\mathcal{S}_{u,f}$. The final skill set is then constructed by combining the selected functional skills with the retrieved user-profile skills:
\begin{equation}
\mathcal{S}_{u}=\mathcal{S}_{u,p}\cup\mathcal{S}_{u,f},
\end{equation}
which is used to augment the recommendation context for downstream recommendation generation.

\subsection{Skill-Guided Trajectory Supervision}

Although the selected skills provide relevant preference evidence and decision strategies, they do not guarantee that the recommendation agent can apply them effectively to candidate-level ranking. CRAFTS therefore introduces
skill-guided trajectory supervision, in which a teacher model generates trajectories demonstrating preference analysis, skill utilization, and candidate comparison to train the recommendation agent
~\cite{hsieh2023distilling}.

For each training instance, we use the instance-specific skill set $\mathcal{S}_u$ constructed in the previous subsection. Given the recommendation context $(\mathcal{H}_u,\mathcal{C}_u)$ and the selected skill set $\mathcal{S}_u$, the teacher model generates a structured recommendation trajectory:
\begin{equation}
Y_u^{T}
=
\left(
R_u^{T},
\widehat{i}_u^{T}
\right),
\end{equation}
where $R_u^{T}$ contains the teacher-generated preference analysis, skill-guided candidate comparison, and ranking rationale, while $\widehat{i}_u^{T}$ denotes the final item selected by the teacher model.

The trajectories generated by the teacher model may contain incorrect decisions or unreliable reasoning. To reduce noisy supervision, we retain only the trajectories whose final predictions match the ground-truth target items. Specifically, we denote each skill-guided trajectory as $\tau_u=(\mathcal{H}_u,\mathcal{C}_u,\mathcal{S}_u,Y_u^{T})$,
and construct the trajectory supervision set as:
\begin{equation}
\mathcal{T}_{\mathrm{distill}}
=
\left\{
\tau_u
\mid
\widehat{i}_u^{T}=i_u^{*}
\right\},
\end{equation}
where $i_u^{*}$ denotes the ground-truth target item for user $u$.

For each retained trajectory, we serialize the structured trajectory $Y_u^{T}$ into a token sequence $y_u^{T}$. The recommendation agent is then trained on these serialized trajectories using an autoregressive sequence modeling objective:
\begin{equation}
\begin{aligned}
\mathcal{L}_{\mathrm{traj}}
=
-
\sum_{\tau_u \in \mathcal{T}_{\mathrm{distill}}}
\sum_{t=1}^{|y_u^{T}|}
\log P_{\theta}
\left(
y_{u,t}^{T}
\mid
\mathcal{H}_u,
\mathcal{C}_u,
\mathcal{S}_u,
y_{u,<t}^{T}
\right),
\end{aligned}
\label{eq:trajectory_loss}
\end{equation}
where $y_{u,t}^{T}$ denotes the $t$-th token in the serialized teacher trajectory, $y_{u,<t}^{T}$ denotes the preceding tokens in the sequence, and $\theta$ denotes the parameters of the recommendation agent. We implement trajectory supervision using LoRA-based parameter-efficient training~\cite{hu2021lora}, where the backbone parameters are frozen and only the low-rank adapter parameters are updated, with a rank of 16, a scaling factor of 32, and a dropout rate of 0.05. 

\subsection{Feedback-Driven Skill Bank Evolution}

Although the initial Skill Bank captures common preference and decision patterns, its coverage may remain insufficient for diverse recommendation scenarios. However, an incorrect prediction does not necessarily indicate missing skill knowledge, as the failure may also arise from imperfect skill retrieval or utilization. Therefore, CRAFTS does not update the Skill Bank from individual failures, but instead analyzes recurring patterns across multiple hard cases before skill evolution.

For each failed recommendation, the system records the user history $\mathcal{H}_u$, candidate set $\mathcal{C}_u$, selected skill set $\mathcal{S}_u$, model prediction $\hat{i}_u$, and ground-truth item $i_u^*$. A failed trajectory is denoted as:
\begin{equation}
e_u =
\left(
\mathcal{H}_u,
\mathcal{C}_u,
\mathcal{S}_u,
\hat{i}_u,
i_u^*
\right),
\end{equation}
where $\hat{i}_u \neq i_u^*$.

CRAFTS accumulates failed recommendation cases and triggers an evolution analysis after collecting five failures. In each round, the LLM jointly analyzes these cases with the current domain-specific Skill Bank $\mathcal{B}^{d}$ to identify recurring failure patterns and potential root causes, determining whether they reveal missing or insufficient transferable skill knowledge:
\begin{equation}
o =
\mathrm{LLM}
\left(
P_{\mathrm{evo}},
\mathcal{E},
\mathcal{B}^{d}
\right),
\end{equation}
where $P_{\mathrm{evo}}$ denotes the skill-evolution prompt, $\mathcal{E}$ denotes the collected batch of failed recommendation cases, and $o$ denotes the resulting analysis and update proposal.

Based on repeated failure evidence, CRAFTS performs two forms of evolution for both user-profile and functional skills. \textit{Create} introduces a new skill when recurring failures reveal an uncovered transferable preference pattern or decision strategy, while \textit{Update} refines an existing skill when the relevant knowledge remains insufficient in preference modeling, applicability conditions, or ranking guidance. Skill updates are proposed only when such deficiencies are supported by recurring failures, thereby reducing unnecessary updates caused by isolated retrieval or utilization errors.

\section{Experiments}

We empirically evaluate CRAFTS by addressing the following research questions (RQs):
\begin{itemize}[leftmargin=*, itemsep=0pt, topsep=2pt]
\item \textbf{RQ1:} How does CRAFTS perform compared with existing methods across different recommendation domains and LLM backbone scales?
\item \textbf{RQ2:} Does skill-guided trajectory supervision enable the recommendation agent to selectively utilize context-relevant transferable skills?
\item \textbf{RQ3:} How do the key components of CRAFTS contribute to the final recommendation performance?
\item \textbf{RQ4:} Does feedback-driven Skill Bank evolution improve recommendation performance by refining the transferable skill space?
\end{itemize}

\subsection{Experimental Setup}

\subsubsection{Datasets.}
We evaluate CRAFTS on three widely used recommendation datasets: LastFM~\cite{cantador2011second}, MovieLens~\cite{harper2015movielens}, and Instruments~\cite{he2016ups}, covering music, movie, and product recommendation scenarios. Since the test set of Instruments is relatively large, 1,000 test instances are sampled to align its scale with LastFM and MovieLens. Dataset statistics are summarized in Table~\ref{tab:dataset_statistics}.

\begin{table}[t]
\centering
\caption{Statistics of the datasets.}
\label{tab:dataset_statistics}
\begin{tabular}{lrrr}
\toprule
Dataset & \#Users & \#Items & \#Interactions \\
\midrule
LastFM & 1,892 & 17,632 & 92,834 \\
MovieLens & 943 & 1,682 & 100,000 \\
Instruments & 339,231 & 84,901 & 500,176 \\
\bottomrule
\end{tabular}
\end{table}

\subsubsection{Evaluation Setting.}
For each test instance, the next interacted item is treated as the positive item given the user's history. Nineteen negative items are randomly sampled from non-interacted items to form a candidate list of 20 items. This protocol is widely adopted in LLM-based recommendation studies~\cite{liao2024llara,cai2025agentic}, and the same candidate lists are used for all methods.

Hit Ratio@1 (HR@1) is adopted as the main metric. A recommendation is counted as a hit if the Top-1 prediction matches the ground-truth item; otherwise, it is counted as a miss.

\subsubsection{Baselines.}
We compare CRAFTS with the following baselines.

\begin{itemize}[leftmargin=*, itemsep=0pt, topsep=2pt]
\item \textbf{AFL.} Agentic Feedback Loop is a recommendation framework that leverages interactive feedback between a recommendation agent and a user agent to adjust subsequent recommendation results~\cite{cai2025agentic}.

\item \textbf{iAgent.} iAgent is a user-side recommendation agent that integrates user instructions, external knowledge, and reranking for candidate optimization~\cite{xu2025iagent}. Its extension, i$^2$Agent, further incorporates dynamic memory mechanisms for personalized user profile modeling.

\item \textbf{MemSkill-Rec.} MemSkill formulates memory operations as learnable and evolvable skills through skill selection, execution, and evolution~\cite{zhang2026memskill}. We adapt MemSkill to recommendation by enabling recommendation skill construction from user histories and interaction trajectories.
\end{itemize}

\subsubsection{Backbone Models.}
To evaluate CRAFTS under different LLM capacities, three Qwen3 models with different parameter scales are adopted as recommendation agent backbones: Qwen3-1.7B, Qwen3-4B, and Qwen3-8B~\cite{yang2025qwen3}. These models represent lightweight, medium-scale, and stronger LLM capacities, respectively. Experiments with these different backbone sizes are conducted to examine whether CRAFTS improves recommendation performance across model scales.

In experiments, the backbone model instantiates the recommendation agent, which performs preference analysis, candidate comparison, and recommendation generation based on user histories, candidate items, and retrieved skills. Qwen3-8B is adopted for teacher trajectory generation, Skill Bank initialization, and feedback-driven Skill Bank evolution. When Qwen3-8B is used as the recommendation-agent backbone, the teacher and student have the same model scale, while the student is further fine-tuned using LoRA on the retained teacher-generated trajectories. For skill retrieval, text representations are obtained using the frozen \texttt{all-MiniLM-L6-v2} encoder.
\subsection{Effectiveness of CRAFTS}
In this section, we evaluate the effectiveness of CRAFTS from three perspectives. First, we compare CRAFTS with representative agent-based and skill-enhanced recommendation methods across different datasets and LLM backbones. Second, we analyze the recommendation agent's responses to user-profile skills with different retrieval relevance before and after trajectory supervision, examining whether trajectory supervision improves its ability to selectively utilize relevant skills. Third, we investigate whether the learned transferable skills can improve recommendation performance for users who are not involved in Skill Bank construction.
\subsubsection{Recommendation Results.}

We compare CRAFTS with representative agent-based and skill-enhanced methods across three datasets and LLM backbones. The results are reported in Table~\ref{tab:recommendation_performance}. Overall, CRAFTS demonstrates competitive performance across different recommendation scenarios and backbone scales.
\begin{table}[t]
\centering
\caption{Recommendation performance (HR@1) across three datasets and three Qwen3 backbones.}
\label{tab:recommendation_performance}

\small
\setlength{\tabcolsep}{4pt}
\renewcommand{\arraystretch}{1.03}

\begin{tabularx}{\columnwidth}{
    >{\raggedright\arraybackslash}p{0.25\columnwidth}
    Y
    Y
    Y
}
\toprule
\textbf{Method}
& \textbf{LastFM}
& \textbf{MovieLens}
& \textbf{Instruments} \\
\midrule

\multicolumn{4}{c}{\textit{Qwen3-1.7B}} \\
\cmidrule{1-4}

AFL
& 0.2136 & 0.4178 & \underline{0.3270} \\
iAgent
& 0.0751 & 0.0679 & 0.1890 \\
i$^2$Agent
& 0.0852 & 0.0827 & 0.1950 \\
MemSkill-Rec
& \underline{0.2552} & \underline{0.4698} & 0.2780 \\

\rowcolor{gray!12}
\textbf{CRAFTS}
& \textbf{0.2669} & \textbf{0.5514} & \textbf{0.3560} \\

\midrule
\multicolumn{4}{c}{\textit{Qwen3-4B}} \\
\cmidrule{1-4}

AFL
& \underline{0.2840} & \underline{0.4973} & 0.3310 \\
iAgent
& 0.1492 & 0.1156 & 0.3320 \\
i$^2$Agent
& 0.1593 & 0.1145 & \underline{0.3430} \\
MemSkill-Rec
& 0.2291 & 0.4867 & 0.2750 \\

\rowcolor{gray!12}
\textbf{CRAFTS}
& \textbf{0.3053} & \textbf{0.5546} & \textbf{0.3600} \\

\midrule
\multicolumn{4}{c}{\textit{Qwen3-8B}} \\
\cmidrule{1-4}

AFL & \underline{0.3202} & \underline{0.5058} & \underline{0.3420} \\
iAgent & 0.2088 & 0.1930 &\ 0.3390 \\
i$^2$Agent & 0.2206 & 0.1410 & \textbf{0.3520} \\
MemSkill-Rec & 0.2509 & 0.3659 & 0.3270 \\

\rowcolor{gray!12}
\textbf{CRAFTS}
& \textbf{0.3335} & \textbf{0.5387} & \underline{0.3420} \\

\bottomrule
\end{tabularx}

\vspace{2pt}
\parbox{\columnwidth}{%
\footnotesize
The best result is shown in \textbf{bold}, and the second-best
result is \underline{underlined}.
}
\end{table}

\begin{table*}[!t]
\centering
\caption{HR@1 of Qwen3-4B with different user-profile skill inputs before and after skill-guided trajectory SFT. The best result in each column is shown in bold.}
\label{tab:user_skill_utilization}
\small
\setlength{\tabcolsep}{9pt}
\renewcommand{\arraystretch}{1.08}

\begin{tabular*}{0.9\textwidth}{
@{\extracolsep{\fill}}
l
cc
cc
cc
}
\toprule
\multirow{2}{*}{\textbf{User-Profile Skill}}
& \multicolumn{2}{c}{\textbf{LastFM}}
& \multicolumn{2}{c}{\textbf{MovieLens}}
& \multicolumn{2}{c}{\textbf{Instruments}} \\
\cmidrule(lr){2-3}
\cmidrule(lr){4-5}
\cmidrule(lr){6-7}

& \textbf{Before SFT} & \textbf{After SFT}
& \textbf{Before SFT} & \textbf{After SFT}
& \textbf{Before SFT} & \textbf{After SFT} \\
\midrule

No User-Profile Skills
& \textbf{0.2419} & 0.2605
& 0.4210 & 0.5323
& \textbf{0.2740} & 0.3160 \\

Retrieved User-Profile Skills
& 0.2243 & \textbf{0.3053}
& \textbf{0.4645} & \textbf{0.5546}
& 0.2530 & \textbf{0.3600} \\

Low-Relevance User-Profile Skills
& 0.2174 & 0.2589
& 0.4634 & 0.5366
& 0.2380 & 0.2990 \\

\bottomrule
\end{tabular*}
\end{table*}

Compared with agent-based baselines, CRAFTS generally achieves higher HR@1 across the evaluated settings. AFL performs competitively on LastFM and MovieLens, while i$^2$Agent obtains strong results on Instruments. However, these methods mainly rely on instance-level feedback or user-specific memory, limiting the reuse of decision knowledge across users and recommendation contexts. In contrast, CRAFTS abstracts reusable preference judgment and candidate comparison patterns into transferable skills and retrieves relevant skills for each recommendation instance, resulting in consistent improvements across domains. Compared with MemSkill-Rec, CRAFTS also achieves clear gains. Although MemSkill-Rec introduces skill-like structures into recommendation, its performance varies across backbone scales. For example, its results decrease on LastFM and Instruments when changing from Qwen3-1.7B to Qwen3-4B, indicating that directly adapting skill frameworks developed for other domains may be insufficient for recommendation. Although such mechanisms can provide useful information for individual recommendation decisions, the accumulated experience remains closely associated with specific users or interaction contexts, making it difficult to directly benefit subsequent recommendation instances involving similar preference patterns. In contrast, CRAFTS learns transferable recommendation skills, retrieves context-relevant skills, and supervises their utilization through skill-guided trajectories, enabling better reuse of decision knowledge for preference analysis and candidate ranking. The relatively smaller gains under Qwen3-8B may be attributed to its stronger inherent reasoning capability, which leaves less room for further improvement from skill-enhanced supervision. Moreover, since Qwen3-8B is also employed as the teacher model, the absence of a capacity gap between the teacher and student may further limit the potential benefits of trajectory supervision.
\begin{figure}[t]
    \centering
    \includegraphics[width=\linewidth]{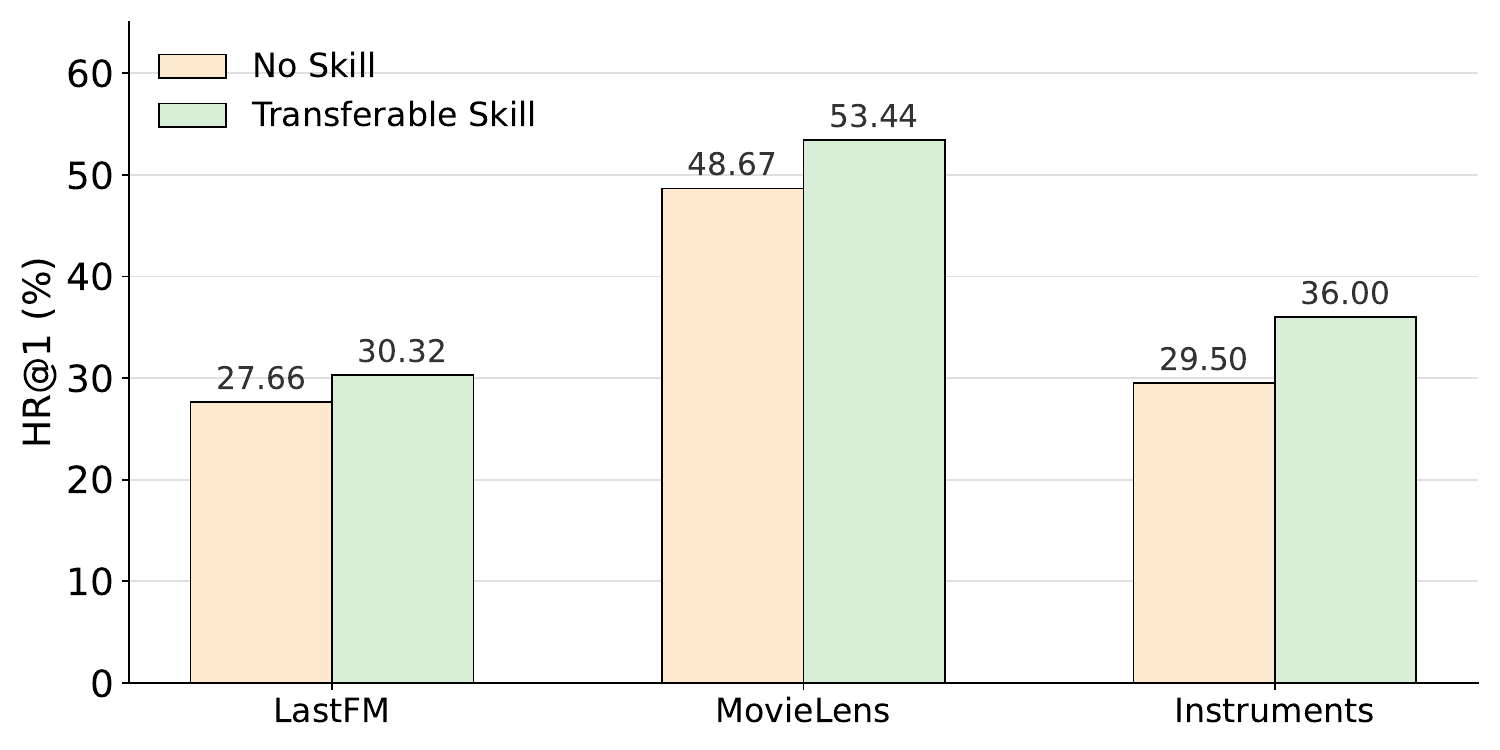}
    \caption{Effectiveness of Skills across Unseen Users.}
    \Description{A comparison of HR@1 with and without transferable skills on LastFM, MovieLens, and Instruments. Using transferable skills learned from other users improves recommendation performance across all three datasets.}
    \label{fig:rs}
    \vspace{-5mm}
\end{figure}

\subsubsection{User-Profile Skill Utilization Analysis.}

Transferable user-profile skills capture shared preference patterns across historical interactions and may not perfectly match every user. Therefore, effective utilization requires the recommendation agent to identify applicable skill signals rather than indiscriminately relying on all retrieved skills. To investigate whether trajectory supervision enables adaptive skill utilization, we compare Qwen3-4B before and after skill-guided trajectory SFT under three settings: \textbf{No User-Profile Skills}, \textbf{Retrieved User-Profile Skills}, and \textbf{Low-Relevance User-Profile Skills}. For each model before and after SFT, the three settings differ only in the user-profile skills provided during inference, with model parameters kept unchanged across settings. In particular, the low-relevance setting replaces retrieved user-profile skills with the same number of low-relevance user-profile skills sampled from the Skill Bank, while keeping functional skills and other recommendation inputs unchanged. The results are reported in Table~\ref{tab:user_skill_utilization}.

Before trajectory supervision, retrieved user-profile skills do not consistently improve performance and may even reduce accuracy, indicating that direct exposure to partially matched skills is insufficient for effective utilization. Since transferable skills are derived from shared preference patterns, some of their signals may not be applicable to the current recommendation instance. Without explicit guidance on skill utilization, the agent may struggle to distinguish useful preference signals from those that are less relevant to the current user. After trajectory supervision, the same retrieved skills consistently outperform the setting without user-profile skills across all datasets. This demonstrates that skill-guided trajectories help the recommendation agent align transferable skill signals with user contexts and selectively apply useful signals for recommendation decisions. Moreover, the improved performance of retrieved skills relative to low-relevance skills after trajectory supervision suggests that the agent becomes more sensitive to the contextual relevance of skill knowledge, rather than merely benefiting from additional skill information.

\subsubsection{Cross-User Skill Transferability Analysis.}
To further investigate whether the learned skills can be transferred beyond the users from which they are derived, we conduct a cross-user skill transferability analysis. Specifically, we partition the users in each dataset into two disjoint subsets, using 80\% of the users for Skill Bank construction and the remaining 20\% for cross-user transferability evaluation. Therefore, the evaluated users rely on transferable skills from the shared Skill Bank rather than user-profile skills constructed from their own interaction histories.

We compare two settings: \textbf{No Skill}, where the recommendation agent only receives user history and candidate items, and \textbf{Transferable Skill}, where the recommendation agent obtains skills from the Skill Bank constructed using other users' interactions. As shown in Figure~\ref{fig:rs}, introducing transferable skills consistently improves recommendation performance across different datasets. This demonstrates that the learned skills capture reusable preference patterns beyond individual users and can provide effective guidance for new recommendation contexts. Although users may differ in their specific interaction histories, they can still share underlying preference patterns and decision characteristics. By abstracting such common patterns into transferable skills, CRAFTS can leverage recommendation knowledge accumulated from other users to complement individual preference evidence, providing additional guidance for understanding latent user interests and distinguishing between candidate items.

\begin{table}[!htbp]
\centering
\caption{Ablation results (HR@1) of CRAFTS with Qwen3-4B.}
\label{tab:ablation}
\small
\setlength{\tabcolsep}{5pt}
\renewcommand{\arraystretch}{1.05}

\begin{tabular}{lcccc}
\toprule
Method & LastFM & MovieLens & Instruments & Drop \\
\midrule
w/o User-Profile Skills
& 0.2824 & 0.4719 & 0.3470 & 9.72\% \\

w/o Functional Skills
& 0.3010 & 0.5461 & 0.3200 & 4.33\% \\

w/o Trainable Retriever
& 0.2808 & 0.5334 & 0.3060 & 8.17\% \\

w/o Trajectory SFT
& 0.2957 & 0.4899 & 0.3220 & 9.21\% \\

w/o Skills in SFT
& 0.2856 & 0.5090 & 0.3240 & 8.30\% \\

w/o Skill Evolution
& 0.2834 & 0.5345 & 0.3150 & 7.13\% \\

\midrule
CRAFTS
& \textbf{0.3053}
& \textbf{0.5546}
& \textbf{0.3600}
& -- \\
\bottomrule
\end{tabular}

\vspace{2pt}
\begin{minipage}{\linewidth}
\footnotesize
Drop denotes the relative decrease in mean HR@1 across the three datasets compared with the full model.
\end{minipage}

\end{table}

\subsection{Ablation Study}

In this section, we investigate the effectiveness of the key components of CRAFTS, including user-profile and functional skills, the trainable skill retriever, skill-guided trajectory supervision, and feedback-driven Skill Bank evolution. All experiments use Qwen3-4B as the backbone model, and the results are reported in Table~\ref{tab:ablation}.

\subsubsection{Impact of User-Profile and Functional Skills.}

To examine the contribution of different skill types, we separately remove user-profile skills and functional skills. User-profile skills provide stable preference evidence, while functional skills provide general strategies for candidate comparison and ranking.

As shown in Table~\ref{tab:ablation}, removing either skill type degrades performance, demonstrating their complementary roles. Removing user-profile skills causes a larger average drop, suggesting that transferable preference patterns provide valuable evidence beyond individual interaction histories. In contrast, removing functional skills has limited impact on LastFM and MovieLens but leads to a larger degradation on Instruments, suggesting that functional skills are particularly useful for fine-grained candidate comparison. By combining transferable preference knowledge with general decision strategies, CRAFTS not only enriches the evidence available for user preference modeling but also provides guidance on how to utilize such evidence in candidate comparison. In this way, the two types of skills work together to support more effective recommendation decisions.

\subsubsection{Impact of the Trainable Skill Retriever.}

To evaluate the necessity of task-specific skill retrieval, we replace the trainable skill retriever with a semantic similarity-based retriever while keeping the Skill Bank, the number of retrieved skills, and the recommendation agent unchanged.

As shown in Table~\ref{tab:ablation}, this variant consistently underperforms the full model, with a relative decrease of 8.17\% in mean HR@1. This performance gap suggests that semantic similarity alone may not adequately reflect the utility of a skill for a specific recommendation task. Although a skill may be semantically related to the user's general interests, its preference signals may not be sufficiently discriminative for ranking the current candidate items. In contrast, the trainable retriever learns task-specific relevance from successful recommendation trajectories, allowing skill selection to account for both the underlying user preferences and the current recommendation context. The results highlight the importance of retrieving skills according to their potential contribution to recommendation decisions rather than semantic similarity alone.

\subsubsection{Impact of Skill-Guided Trajectory Supervision}

To disentangle the effects of general trajectory supervision and skill-specific supervision, we consider two ablation variants. The first (w/o Trajectory SFT) removes trajectory supervision entirely. The second (w/o Skills in SFT) retains trajectory supervision but excludes transferable skills from both teacher trajectory generation and student fine-tuning. The resulting trajectories provide general supervision for preference analysis and candidate ranking without explicit guidance on skill utilization. During inference, both variants retain the same skill retrieval and selection mechanism as CRAFTS.

As reported in Table~\ref{tab:ablation}, removing trajectory supervision results in a relative decrease of 9.21\% in mean HR@1. General trajectory supervision improves recommendation performance compared with the variant without SFT, while skill-guided trajectory supervision yields further improvements, highlighting the importance of explicit skill-utilization guidance.

Since retrieved skills summarize shared patterns across users, they may only partially match the current user and candidate context. While general recommendation trajectories provide supervision for preference analysis and candidate ranking, skill-guided trajectories additionally demonstrate how relevant skill signals should be identified and integrated into these processes. This explicit guidance enables the recommendation agent to adapt transferable skills to individual recommendation contexts rather than merely improve its general ranking ability.

\subsubsection{Impact of Feedback-Driven Skill Bank Evolution}

To evaluate feedback-driven Skill Bank evolution, we disable refinement based on failed recommendation trajectories while keeping the initial Skill Bank, skill retriever, and trajectory supervision unchanged.

As shown in Table~\ref{tab:ablation}, removing Skill Bank evolution decreases recommendation performance, demonstrating the benefit of refining transferable skills with failure signals. Although the initial Skill Bank captures recurring preference and decision patterns, certain preference patterns or candidate comparison strategies may remain underrepresented in diverse recommendation scenarios. Failed recommendation trajectories complement the initial knowledge by revealing limitations that emerge when skills are applied to specific recommendation instances. By incorporating recurring failure patterns into skill refinement, CRAFTS can progressively improve the stored recommendation knowledge according to actual recommendation outcomes. The observed performance gains highlight the value of maintaining an evolving rather than static Skill Bank.

\begin{figure*}[!t]
    \centering
    \includegraphics[width=\textwidth]{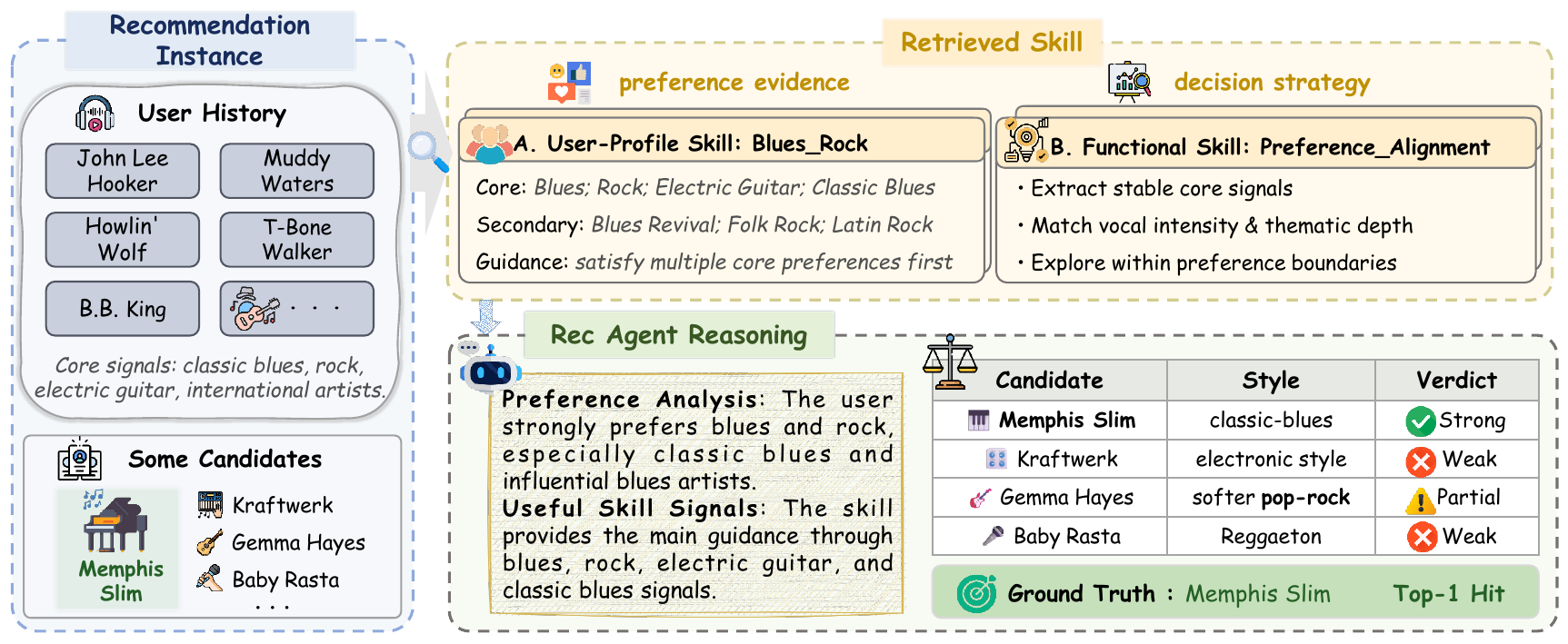}
    \caption{A LastFM case illustrating how CRAFTS uses user-profile and functional skills to guide preference analysis and candidate comparison. Only a subset of the 20 candidates is shown.}
    \Description{A LastFM recommendation example in which CRAFTS retrieves the Blues Rock user-profile skill and the Preference Alignment functional skill. These skills guide preference analysis and candidate comparison, leading to the selection of Memphis Slim, which matches the ground-truth item.}
    \label{fig:case_study}
\end{figure*}
\subsection{Case Study}

To illustrate how CRAFTS leverages retrieved skills for recommendation, we present a case from the LastFM dataset in Figure~\ref{fig:case_study}. Based on the user's historical interactions, the LLM identifies a stable preference for blues and rock, particularly classic blues and electric guitar. Given the current recommendation context, CRAFTS selects relevant skills from the Skill Bank, including the user-profile skill \texttt{Blues\_Rock} and the functional skill \texttt{Preference\_Alignment}, which provide preference evidence and decision strategies, respectively. During candidate comparison, Gemma Hayes exhibits some rock characteristics but only partially matches the user's preference for classic blues, while Kraftwerk and Baby Rasta are more associated with electronic and reggaeton styles and thus deviate from the user's core interests. In contrast, Memphis Slim shows a strong alignment with key signals such as classic blues and is therefore identified as the most suitable candidate, consistent with the ground-truth item.

Overall, this case demonstrates the complementary roles of the two skill types in CRAFTS. User-profile skills help the recommendation agent identify stable user interests, while functional skills provide candidate-level comparison guidance based on these preference signals. This qualitative observation is consistent with the ablation results, which show that both types of skills contribute to recommendation performance. This case illustrates how CRAFTS translates retrieved skills into decision criteria for distinguishing candidates with different characteristics and identifying the ground-truth item.

\section{Conclusion and Future Work}
In this paper, we propose CRAFTS, a framework for learning transferable recommendation skills for agentic recommender systems. CRAFTS abstracts reusable preference judgment and candidate comparison patterns from historical interactions, organizes them into an evolvable Skill Bank, and employs skill-guided trajectory supervision to enable adaptive skill utilization for recommendation decisions. Failed recommendation trajectories are further leveraged to refine and expand the transferable skill space. Extensive experiments on multiple benchmarks demonstrate the effectiveness of CRAFTS compared with representative agent-based and skill-enhanced methods, validating the value of transferable skill learning. Future work will explore cross-domain skill transfer and flexible skill composition to further improve the generalization and reuse of recommendation knowledge.

\section{Ethical Considerations}
CRAFTS maintains a shared and evolving Skill Bank derived from historical user interactions, which may raise privacy and safety concerns. To reduce the risk of exposing individual preference information, CRAFTS abstracts user histories into high-level preference patterns without explicitly retaining user-identifying information or specific interacted items. Beyond privacy protection, the safety of shared skill knowledge also warrants attention. Recent research has shown that internal beliefs held by LLM-based agents can lead to biased behaviors~\cite{wang2026agents}. Similarly, biased or unreliable decision patterns may be introduced during skill construction and reinforced through subsequent skill evolution, highlighting the need for explicit safety validation before cross-user skill reuse.

Even when shared skills capture valid preference patterns, their applicability may vary across individual users. Excessive reliance on such skills may overlook niche user interests, reinforce existing preferences, and consequently reduce recommendation diversity. CRAFTS employs relevance-based skill retrieval and skill-guided trajectory supervision to encourage selective skill utilization according to individual recommendation contexts. However, these mechanisms primarily optimize recommendation accuracy without explicitly considering fairness and diversity objectives. Future work will explore privacy-preserving skill construction, safety-aware skill evolution, and fairness- and diversity-aware skill utilization to promote responsible knowledge transfer in agentic recommender systems.

\begin{acks}
This work is partially supported by the National Key Research
and Development Program of China (Grant No. 2024YFC3014900),
the National Foreign Experts Program of China (Grant No. H20251055),
the Chongqing University Spark Program (Grant No. 2026CQUXH-B18),
and the Australian Research Council (Grant No. DE250100613).
\end{acks}

\newpage

\bibliographystyle{ACM-Reference-Format}
\bibliography{sample-base}

\appendix
\raggedbottom

\section{Representative Skills and Generation Prompts}
\label{app:skill_examples}

To provide a concrete view of the transferable knowledge stored in
CRAFTS, we present representative user-profile and functional skills
from the MovieLens Skill Bank. User-profile skills are first extracted
from individual interaction histories and then aggregated into
group-level preference patterns, whereas functional skills encode
reusable candidate-ranking strategies. Both types of skills can be
further refined through feedback-driven Skill Bank evolution.
\vspace{-1mm}
\subsection{Representative Skill Examples}
\vspace{-1mm}
\begin{skillbox}{User-Profile Skill: Character-driven Drama with Cultural Context}
\small

This skill represents a reusable preference pattern shared by users
whose interaction histories exhibit consistent interests in
character-centered and thematically rich films.

\vspace{-1mm}
\begin{itemize}[leftmargin=*, itemsep=1pt, topsep=2pt, parsep=0pt]
    \item \textbf{Core Preferences:}
    Character-driven drama; cultural or historical context; mature
    themes; emotional depth.
    
    \item \textbf{Secondary Preferences:}
    Narrative complexity; mature thematic exploration; subtle suspense
    elements; cinematic visual style.
    
    \item \textbf{Exploration Boundary:}
    Stay close to the dominant signals of character-driven drama and
    cultural or historical context.
    
    \item \textbf{Ranking Guidance:}
    First satisfy multiple core preferences; use secondary preferences
    for tie-breaking.
\end{itemize}

\end{skillbox}

\vspace{-1mm}

\begin{skillbox}{Functional Skill: Preference Alignment}
\small

This skill specifies a general candidate-comparison strategy for
aligning recommendations with stable preference evidence.

\vspace{-1mm}
\begin{itemize}[leftmargin=*, itemsep=1pt, topsep=2pt, parsep=0pt]
    \item \textbf{Purpose:}
    Keep recommendations aligned with strong recurring preference signals.
    
    \item \textbf{When to Use:}
    Default recommendation scenarios or when the interaction history
    exhibits stable preference patterns.
    
    \item \textbf{Input Signals:}
    History keywords; repeated creators or genres; consistent likes.
    
    \item \textbf{How to Apply:}
    (1) Identify the 2--3 strongest recurring preference cues;
    (2) rank candidates satisfying multiple strong cues above one-off
    matches; and
    (3) use the prior recommendation only as supporting evidence rather
    than the sole decision rule.
    
    \item \textbf{Constraints:}
    Do not overfit to a single token; do not ignore explicit dislikes.
\end{itemize}

\end{skillbox}

\vspace{-1mm}
\subsection{Skill Generation Prompts}
\label{app:skill_prompts}

Due to space limitations, we present condensed versions of the prompts
used in CRAFTS. We retain the core task definition, input/output format,
and major generation constraints, while omitting repetitive
dataset-specific descriptions and implementation details.

\vspace{1mm}

\begin{promptbox}{(1) Initial Skill Construction}
\footnotesize

User-profile skills are first constructed from users' historical
interactions. Given an interaction history, the LLM is instructed to
abstract stable and reusable preference patterns rather than directly
copying interacted items. The prompt is summarized as follows:

\vspace{1mm}
\noindent\textbf{System.}
You are an expert recommender-system profile designer. Infer a reusable
and discriminative user-profile skill from a user's interaction history.
Summarize latent preferences rather than copying exact item names.
The resulting profile should provide transferable preference evidence
for subsequent candidate ranking.

\vspace{1mm}
\noindent\textbf{Input.}
User interaction history: \texttt{\{history\}}.

\vspace{1mm}
\noindent\textbf{Output.}
Return valid JSON containing
\texttt{name}, \texttt{core\_preference}, \texttt{secondary\_preference}, 
\texttt{exploration\_boundary} and \texttt{ranking\_guidance}.

\vspace{1mm}
\noindent\textbf{Requirements.}
Infer high-level preference patterns from the interaction history;
distinguish stable core preferences from weaker tendencies;
specify acceptable exploration boundaries and compatibility
constraints; and provide explicit guidance for candidate ranking.
The generated skill should be reusable across similar recommendation
contexts rather than being tied to individual items.

\vspace{1mm}
Semantically similar user-profile skills are subsequently aggregated
to transform instance-level preference descriptions into more stable
group-level preference patterns. In parallel, the LLM generates a
compact set of functional skills according to the recommendation task
and general recommendation principles. These functional skills encode
reusable candidate-comparison strategies, with each skill specifying
its purpose, applicable conditions, input signals, application rules,
and constraints.

\end{promptbox}

\vspace{-1mm}

\begin{promptbox}{(2) Feedback-Driven Skill Bank Evolution}
\footnotesize

The initial Skill Bank may still provide insufficient coverage for
diverse recommendation scenarios. CRAFTS therefore uses failed
recommendation trajectories as feedback to jointly refine user-profile
and functional skills. Specifically, the LLM receives the current Skill
Bank together with a batch of failed cases and identifies recurring
failure patterns and missing or insufficient decision knowledge.
The evolution prompt is summarized as follows:

\vspace{1mm}
\noindent\textbf{System.}
You are an expert skill-bank designer for a recommender system.
Analyze failed recommendation cases and propose evolutions to the
Skill Bank. You may update existing skills or create new ones, but
every proposed skill must be reusable across users. Return valid
JSON only.

\vspace{1mm}
\noindent\textbf{Input.}
Dataset scope; existing functional skills; existing user-profile skills;
and failed recommendation cases. Each failed case contains the user
history, candidate set, final recommendation, ground-truth item,
user feedback, selected skills, preference query, state summary,
and recent interaction history.

\vspace{1mm}
\noindent\textbf{Output.}
Return JSON containing
\texttt{analysis},
\texttt{functional\_skill\_updates}, and
\texttt{user\_skill\_updates}.
For each proposed update, specify whether it \texttt{updates} an
existing skill or \texttt{creates} a new skill.

\vspace{1mm}
\noindent\textbf{Requirements.}
Prefer refining existing reusable skills over creating one-off skills;
do not construct skills tied to exact items; functional skills should
specify routing conditions and application rules, while user-profile
skills should capture transferable preference patterns. At most one
functional-skill update and one user-profile-skill update are produced
in each evolution step; otherwise, the corresponding update list is
left empty.

\end{promptbox}

\end{document}